\documentclass[preprint,showpacs,preprintnumbers,amsmath,amssymb,superscriptaddress]{revtex4}

\usepackage{graphicx}
\usepackage{dcolumn}
\usepackage{bm}

\begin{document}
\preprint{APS/123-QED}

\title{Topological Defects in Gravitational Lensing Shear Fields}
\author{Vincenzo Vitelli, Bhuvnesh Jain, Randall D. Kamien}
\affiliation{Department of Physics and Astronomy, University of Pennsylvania, Philadelphia PA, 19104}

\newcommand{\vecr}{{\bf r}}
\newcommand{\vecx}{{\bf x}}
\newcommand{\veck}{{\bf k}}
\newcommand{\vectheta}{{\bm \theta}}
\newcommand{\vecl}{{\bm \ell}}
\newcommand{\simgt}{\lower.5ex\hbox{$\; \buildrel > \over \sim \;$}}
\newcommand{\simlt}{\lower.5ex\hbox{$\; \buildrel < \over \sim \;$}}

\begin{abstract}
Shear fields due to weak gravitational lensing have characteristic
coherent patterns. We describe the topological defects in shear fields
in terms of the curvature of the surface described by the lensing
potential. A simple interpretation of the characteristic defects is
given in terms of the the umbilical points of the potential surface 
produced by ellipsoidal halos. We show simulated lensing shear 
maps and point out the typical defect configurations. Finally, we show 
how statistical properties such as the 
abundance of defects can be expressed in terms of the correlation function  
of the lensing potential. 
\end{abstract}

\pacs{Valid PACS appear here}
\maketitle

\section{Introduction}

The central tenet of Einstein's General Relativity is that massive bodies curve space-time. 
As a result, light from distant galaxies is deflected by mass distributions
encountered along the line of sight. The images of distant galaxies, that act as sources, are magnified and sheared -- this effect is known as gravitational
lensing. 
The most striking manifestation of gravitational lensing effects, know as strong lensing, consists in the formation of multiple images of a $single$
background galaxy. In the special case in which a source, a very large mass and the observer happen to be approximately aligned an Einstein ring can be observed and from its diameter the lensing mass can be inferred \cite{Narayan}.

Along typical lines of sight the lensing effect is weak,
leading to percent level magnifications and shears. 
However, such a small signal can still be detected from a $statistical$ analysis
that relies on the coherence of the shear field over the sky. This 
allows us to infer how the dark matter is concentrated around galaxies 
and galaxy clusters, as well as providing a testing ground for dark energy 
and modified gravity theories \cite{Mellier,HoekstraJain}.

In this paper we explore the connection between the theory of
topological defects and the spatial patterns of shear fields due to
weak gravitational lensing. The starting point of our approach rests on
an analogy between gravitational lensing shear fields, as a probe of structure formation on cosmological scales, and
the anisotropic optical or mechanical response of materials, as a probe of their inhomogeneous structure on microscopic scales.
As an illustration, the topological defects in the local shear field of 
an elastic medium reflect
the external deformations applied to the solid \cite{Irvine}. Similarly for thin liquid crystal films confined on a curved substrate, the density 
of topological defects depends on the inhomogeneous curvature of the underlying surface \cite{geomgenerate,Fernandez-Nieves07,Santangelo07}. In a somewhat 
different context, the distribution of  optical singularities can be 
used to shed light on the 
statistics of randomly polarized light fields \cite{Mark1}. Here we suggest that 
topological defects in the cosmic shear can be used as a 
probe of the gravitational potential generated by the lensing mass fluctuations on large scales.

The outline of this work is as follows. In \S 2, the basic 
formalism of weak gravitational lensing and the criteria to 
identify topological defects in shear fields are presented. Our geometric approach is presented in
\S 3 where the defects in the shear field are related to the properties of an imaginary surface whose lines of constant height are defined by the contour lines 
of the underlying gravitational potential responsible for the lensing. This mapping is applied to describe the
characteristic behavior of ellipsoidal structures in the mass
distribution and to more realistic shear lensing maps obtained using N-body computer simulations. 
In \S 4 we turn to the case of a random mass distribution whose gravitational potential is 
a Gaussian variable that describes the stochastic geometry of a surface. This assumption (valid on large angular scales)
allows us to express the 
defect density in terms of the 
two point correlation function of the gravitational potential.
The latter is estimated for 
the standard Cold Dark Matter model for large-scale structure in
the universe and compared to the results of simulations. 
Corrections for weakly non-Gaussian fields can be calculated using 
perturbation theory as discussed in \S 5 where ideas for further work
are briefly sketched. 

\section{Topological Defects in Lensing Shear Fields}

In this section, we review the
basic formalism of gravitational lensing that relates the deformation
of the shapes of background galaxy images to the projected mass density
responsible for lensing.  

\subsection{Basics of Gravitational Lensing}

Consider a source whose true angular position on the sky makes an
angle $\beta$ with an arbitrary optic axis. As a result of the
lensing, an observer sees the light ray as coming from an image at an
angle $\theta$ (from the optic axis) that differs from $\beta$ by the
(reduced) deflection angle $\alpha$. These angular displacements can
also be viewed as two dimensional vectors
$\{\vec{\alpha},\vec{\beta},\vec{\theta}\}$ on a locally flat sky (see
\cite{Bartelmann:1999yn} for a review). These vectors are related
by: ${\vec \beta} = {\vec \theta} - {\vec \alpha}$. 

\begin{figure}
\includegraphics[width=0.5\textwidth]{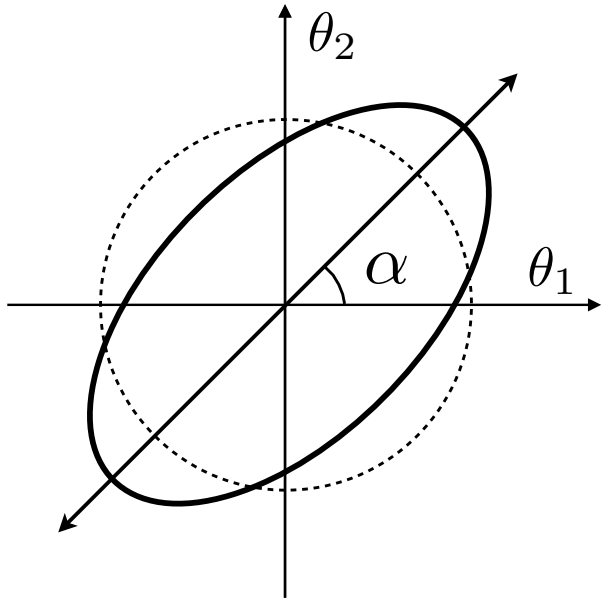}
\caption{\label{fig:ellipse1} The light from a distant source of circular shape (drawn as a dashed line) is deformed into an ellipse (continuous line). The local direction of the major axis of the ellipse can be represented by a double headed vector that forms an angle $\alpha$ with the $\hat{x}$ axis of a cartesian coordinate system. The positions in a small patch of the sky are measured by the angles $\{\theta_1, \theta_2\}$ within the flat sky approximation.}
\end{figure}

In this notation, the image distortion caused by gravitational lensing is
locally described by $A_{ij}$, the Jacobian matrix of the
transformation between $\vec{\beta}$ and $\vec{\theta}$ which reads 
\begin{eqnarray} 
A_{ij} \equiv \frac{\partial \beta_i}{\partial \theta_j}=\delta_{ij}- \frac{\partial \alpha_{i}(\vec{\theta)}}{\partial \theta_j}=\delta_{ij}- \frac{\partial^2 \psi(\vec{\theta})}{\partial \theta_i \partial \theta_j} 
\label{eq:1}
\end{eqnarray}
where we used the relation $\vec{\alpha}=\nabla_{\vec{\theta}}$ $\psi$
between the deflection angle and the gradient of the projected
(two-dimensional) Newtonian potential $\psi(\vec{\theta})$. The latter
satisfies the Poisson equation  
\begin{eqnarray}
\frac{1}{2}(\psi_{11}+\psi_{22}) \equiv \frac{1}{2}\nabla^{2}_{\theta} \psi = \kappa(\vec{\theta})
\label{eq:2}
\end{eqnarray}
where the convergence, $\kappa(\vec{\theta})$, is given by the
weighted projection of the mass density fluctuation field, see
Ref. \cite{Dodelson:2003} for precise definitions. Equation
(\ref{eq:2}) fixes the trace of the deformation matrix $A_{ij}$. The
other two independent components of $A_{ij}$ can be rewritten in terms
of a shear tensor ${\bf \gamma}=\{\gamma_{1},\gamma_{2}\}$ and 
the angle $\alpha$ as
\begin{eqnarray}
\gamma_{1}(\vec{\theta})&=&\frac{1}{2}(\psi_{11}-\psi_{22})\equiv
\gamma(\vec{\theta}) \cos\left[2\alpha(\vec{\theta})\right] \nonumber \\ 
\gamma_{2}(\vec{\theta})&=&\psi_{12}=\psi_{21}\equiv \gamma(\vec{\theta})  \sin\left[2\alpha(\vec{\theta}) \right]  
\label{eq:3}
\end{eqnarray}
where $\gamma=\left(\gamma_1 ^2 + \gamma_2 ^2\right)^{1/2}\ge 0$ and
$\alpha$ is the orientation angle of $\gamma$. The
physical interpretation of the shear tensor 
is straightforward once the components of the Hessian $\psi_{ij}$ are expressed in terms of $\gamma_1$, $\gamma_2$ and $\kappa$ with the aid of Equations (\ref{eq:2}-\ref{eq:3}) and then substituted into Eq. (\ref{eq:1}).  The result reads
\begin{eqnarray}
A_{ij}=(1-\kappa)\begin{pmatrix}
1 & 0 \\
0 & 1 
\end{pmatrix}-\gamma \begin{pmatrix}
\cos2\alpha & \sin2\alpha  \\
\sin2\alpha & -\cos2\alpha 
\end{pmatrix}
\label{eq:4}
\end{eqnarray}
The eigenvalues of this matrix are $\{1-\kappa-\gamma,1-\kappa+\gamma\}$ with eigenvectors $v_1=[\cos\alpha,\sin\alpha]$ and $v_2=[-\sin\alpha,\cos\alpha]$, respectively.  The direction of shear is $v_1$:
when the deformation tensor $(A^{-1})_{ij}$ acts on a test circular source,
it magnifies the image isotropically, by the first term in
Eq. (\ref{eq:4}), and it deforms it into an ellipse with major axis of magnitude $(1-\kappa-\gamma)^{-1}$ oriented
in the $\alpha$ direction, by the second term (see Fig. \ref{fig:ellipse1}). 

In the sub-section below we will consider the pattern of the shear on
the sky induced by the spatial variation of $\psi({\vec\theta})$. Over
patches of order a degree on a side, characteristic tangential
patterns around lensing galaxy clusters and galaxy groups are
observable. 

Thus measuring the
anisotropy introduced in the light distribution from distant galaxies
allows us to infer information about mass
inhomogeneities along the line of sight that generated the lensing
shear.  

\subsection{Topological defects}

The local orientation of the shear can be visualized by means of a
line field, that is to say a vector $\bf{n}$ with both ends identified
$\bf{n}=-\bf{n}$. This is analogous to the description of interacting liquid crystal molecules in
condensed matter physics, where the director field $\bf{n}$ denotes the local orientation of the molecules. In that context, it is often useful to
develop a coarse grained view of the many-body system by concentrating
on {\it global} configurations of the director field called disclinations \cite{nelsonbook}. Disclinations are topological defects like
vortices in fluid mechanics that can be classified according to the index of the streamlines around them. 

\begin{figure}
\includegraphics[width=0.5\textwidth]{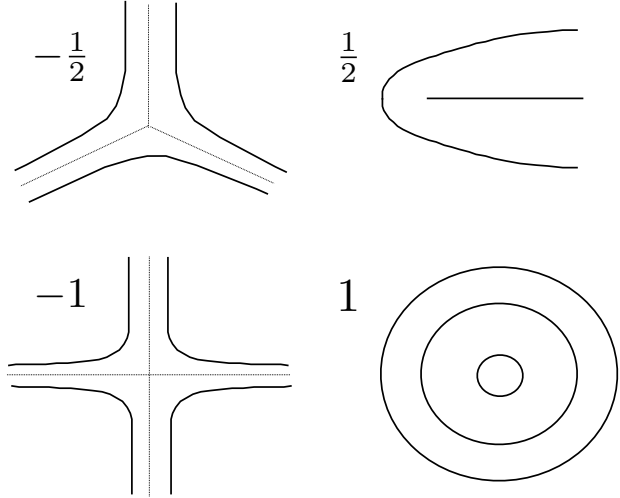}
\caption{\label{fig:defects} (Color Online) Basic classification of topological defects labeled by their respective index. The index $m$ is found by evaluating a line integral of $\nabla \alpha$ around any contour surrounding the singularity at the core of the topological defect and dividing by $2 \pi$, see Equations (\ref{eq:1}) and (\ref{eq:2}).}
\end{figure}

For example , the shear field angle $\alpha(\theta_{1},\theta_{2})$ has a topological defect of index $m$ if
\begin{equation}
\oint \nabla \alpha(\theta_1,\theta_2) \cdot dl=2 \pi m
\label{eq:5}
\end{equation}
around any contour that surrounds the location of the defect ${\vec{\theta}_d}$. Upon using Stokes theorem, Eq. (\ref{eq:5}) can be cast in the differential form 
\begin{equation}
\nabla \times \nabla \alpha(\theta_1,\theta_2) = 2 \pi m \!\!\!\! \quad \delta_{D}^2(\vec{\theta} - \vec{\theta_d})
\label{eq:5bis}
\end{equation}
where $\delta_{D}^2(\vec{\theta} - \vec{\theta_d})$ denotes the Dirac delta function at the location of the defect.
As a result, topological defects are analogous to localized regions of quantized magnetic flux with charge proportional to the winding number $m$, and $\nabla \alpha$ plays the role
of the electromagnetic gauge field. The (shear) field lines corresponding to
topological defects of different indexes are reproduced schematically
in Fig. \ref{fig:defects}.  The symmetry of the shear field requires that the defect index $m$ must be a half-integer.
The negatively
charged defects correspond to the case in which the shear field
changes in the clockwise direction while travelling counter-clockwise
around the contour encircling the defect. 

Global rotations by a constant angle at every point in the flat sky,
$\alpha(\vec{\theta}) \rightarrow \alpha(\vec{\theta}) + c$, do not
change the index of a topological defect since the definition in
Eq. (\ref{eq:4}) depends only on the gradient of $\alpha$. For
example, the vortex-like defect in Fig. \ref{fig:defects} has the
same $m=1$ index as a sink which is obtained from the vortex by a
$c=\pi/2$ rotation. This invariance distinguishes the topological
classification adopted in this work from the more common E-B modes
classification adopted in previous studies of the CMB polarization
field and of gravitational lensing. For example a 45 degree rotation of each shear around an $m=1$ defect does not change its topological charge, but it changes an E mode to a B mode. In this study, we concentrate on
global topological features of the shear field whose positions on the
sky are related (non-locally) to 
the large scale mass fluctuations responsible for the lensing.     

\section{The induced geometry of the lensing shear}

In this section, we derive the connection that exists between the topological defects in the shear field and the projected gravitational potential $\psi(\vec{\theta})$ generated by the mass fluctuations. It is convenient to view $ \psi(\theta_1,\theta_2)$ as the height function of a non-intersecting two dimensional surface.  In the weak lensing regime, the derivatives of the potential $\psi$ are small; the differential geometry of the surface is then approximated by the properties of the Hessian $\psi_{ij}$ \cite{Kamien}.  We show that the shear direction corresponds to the principal direction of maximal curvature on the surface, provided that $\vert\kappa\vert,\gamma\ll 1$.  

\subsection{Defects as umbilical points}

On such a two dimensional surface, we can step away from any given point in an infinite number of directions so that for each direction a curvature
is defined. As the direction is smoothly varied,  two perpendicular directions of {\it principal curvatures} can be found for which the curvature is maximal and minimal.
The two principal curvatures, henceforth denoted by $\{\kappa_1,\kappa_2\}$, are the eigenvalues of the Hessian matrix $\psi_{ij}$:
\begin{equation}
\{\kappa_{1},\kappa_2\} = \frac{1}{2}\left(\psi_{11}+\psi_{22} \pm {\sqrt{4\psi_{12}^2 +(\psi_{11}-\psi_{22}}})^2\right) =\kappa \pm \gamma
\label{eq:5bis}
\end{equation}
The eigenvectors give the local directions of the principal axes. Since the magnitude of the shear field $\gamma$ is positive, $\kappa_{1}=\kappa + \gamma$ corresponds to  
the principal direction of maximal curvature. The Hessian matrix $\psi_{ij}$ determines the deformation matrix $A_{ij}$ according to Eq. (\ref{eq:1}). Upon comparing Eq. (\ref{eq:5bis}) with Eq. (\ref{eq:4}) and the discussion following it, we can conclude that the direction of maximal curvature is the shear direction along which a reference circular source is stretched most by the lensing potential (eg. it points along the major axis of the ellipse in Fig. \ref{fig:ellipse1}).

The two principal curvatures are equal when $4\psi_{12}^2
+(\psi_{11}-\psi_{22})^2$ is equal to zero. This defines the umbilical
points of the surface: points where the principal directions are
undefined and the surface is locally spherical or flat \cite{Hilbert}. 

We can now connect the geometry of the surface
which represents the 
gravitational potential and the topological defects in the lensing shear field. 
At points where the shear field $\gamma$ has a
topological defect, both of its 
components, see Eq. (\ref{eq:3}) must vanish, because its local
direction is undetermined. {\it This ensures that the two principal
curvatures are equal. As a result, topological defects of the shear
field can be identified as the umbilical points 
of the induced surface. } Indeed the umbilical points can be
classified according to the index of the
principal direction vector field, which is either $+\frac{1}{2}$ 
or $-\frac{1}{2}$  at an umbilical point. As 
we shall see this corresponds to topological defects of index $\pm
\frac{1}{2}$ in the shear field. 

The local angle $\tilde{\alpha}(\theta_1,\theta_2)$ by which the
coordinate axis needs  to be rotated to overlap with the principal
axes (given by the eigenvectors of the Hessian of the height function
$\psi$) is
\begin{equation}
\tan 2 \tilde{\alpha}= \frac{2 \psi_{12}}{\psi_{11}-\psi_{22}}
\label{eq:6bis}
\end{equation}

A comparison of Eq.~(\ref{eq:6bis}) with the definition of the shear
components $\gamma_1$ and $\gamma_2$ in Eq.~(\ref{eq:3}) shows
that $\tilde{\alpha} = \alpha$, the angle that specifies the
orientation of the shear $\gamma$. Thus the principal direction of maximal curvature on the surface
tracks the shear field.

In the next sections we will show how this mathematical mapping leads
to connections between the defect locations and the
distribution of mass that generates 
the gravitational potential. In addition the identification between
topological defects and umbilical points of a random surface will allow
us to relate the density of
defects to statistical quantities such as the two-point correlation
function of the projected gravitational field.

In practice, one would attempt to identify defects either in 
direct experimental measurements of shear using galaxy images, or ray tracing
computer simulations that provide the shear field on a discrete grid. 
In both cases, it is necessary to consider how to interpret the notion
of a defect (introduced via Eq. (\ref{eq:5}) in the continuum limit)
on a square grid. We have carried out such a measurement using the net 
change in the angle $\alpha$ around closed loops at each pixel vertex, 
using techniques
similar to those of \cite{Huterer}. Further details of the numerical
aspects will be presented elsewhere. 

\subsection{The ellipticity of haloes}

To illustrate the formal ideas presented in the last section consider
the simple case of an elliptical potential field $\psi(x,y)$ generated, for example,
by an $isolated$ distribution of mass which is not axisymmetric, namely an
elliptical halo. (Note, however, that even if the mass distribution is locally axisymmetric the
corresponding potential can be perturbed into an elliptical shape by the tidal field of nearby
objects, see next section \cite{Schneider}.).

For simplicity, consider two simple models of elliptical haloes whose gravitational potentials are given in dimensionless variables by 
\begin{figure}
\includegraphics[width=1\textwidth]{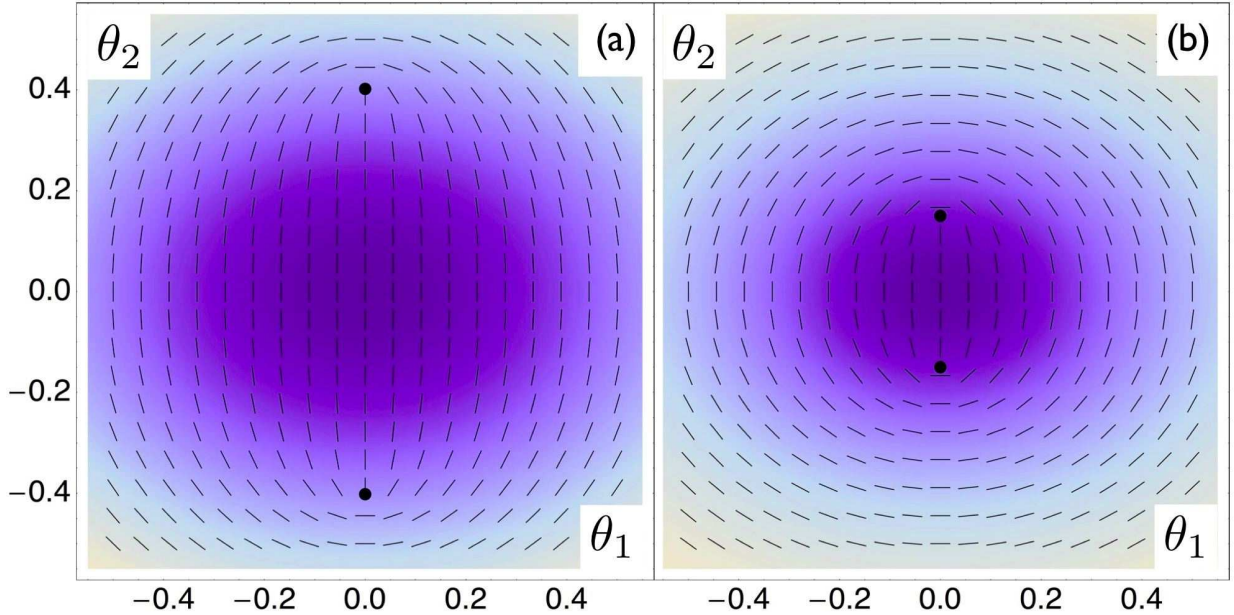}
\caption{\label{fig:ellipse} Panel (a) and (b) show contour plots of the gravitational potentials for the elliptical halos in Eq. (\ref{eq:7}) and Eq. (\ref{eq:7b}) respectively (dark blue is most negative). The parameters were set to $\epsilon=0.3$ and $c=0.2$. The two black dots indicate the umbilical points
that correspond to the two $+1/2$ topological defects in the shear field. The wiskers represent the local direction of the cosmic shear which is oriented along the
principal direction with maximum curvature.}
\end{figure}
\begin{eqnarray}
\psi (\theta_1,\theta_2)= -\exp{\left[-\left(\frac{\theta_1 ^2}{1+\epsilon}+ \frac{\theta_2 ^2}{1-\epsilon}\right)\right]} 
\label{eq:7}
\end{eqnarray}
and  
\begin{eqnarray}
\psi (\theta_1,\theta_2)= -\frac{1}{\sqrt{\frac{\theta_1 ^2}{1+\epsilon}+ \frac{\theta_2 ^2}{1-\epsilon}+c}} .
\label{eq:7b}
\end{eqnarray}
Figures \ref{fig:ellipse}(a) and \ref{fig:ellipse}(b) show the contours plots of the two potentials in Eqs. (\ref{eq:7}) and (\ref{eq:7b}) respectively, overlayed onto the
corresponding (normalized) shear fields represented by wiskers. The umbilical points are labelled by black
dots. Comparison of the shear field around each
defect in this figure with the defects shown in 
Fig. \ref{fig:defects} allows a clear identification  
of the two umbilical points with two $+ 1/2$ topological defects. The line joining the two defects is perpendicular to the major axis of the ellipsoidal contours of the potential (and of the mass). This follows from the fact that the shear field is oriented along the principal direction of maximal curvature. The defects
separation, $s$, measured in the same units of the angular distances $\{\theta_1,\theta_2\}$ adopted for the potential in Eq. (\ref{eq:7}), reads
\begin{equation}
s= 2 \sqrt{\frac{(1-\epsilon) \epsilon}{1+\epsilon}} .
\label{eq:8}
\end{equation}
while the separation for the potential in Eq. (\ref{eq:7b}) reads
\begin{equation}
s= 2 \sqrt{\frac{2 c (1-\epsilon) \epsilon}{3+\epsilon}} .
\label{eq:8b}
\end{equation}
In the limit of zero ellipticity, $\epsilon=0$ and the separation between the two defects, $s\sim\sqrt{\epsilon}$, goes to zero. The two $+1/2$ defects coalesce into a single
vortex of index $+1$ which is expected from an isolated 
axisymmetric mass distribution \cite{Bartelmann:1999yn}. Qualitatively similar behavior is expected for ellipsoidal haloes
with different potential profiles. We have also checked this for a non-singular logarithmic potential. 

Hence, the
separation of the two nearby defects is a measure of the ellipticity
of the haloes. This is potentially useful 
since large-scale structure studies have shown that the mass distribution in
the universe can be well approximated (especially for statistical
purposes) as a network of elliptical halos of varying mass and
concentration \cite{ShethTormen2002,Cooray}. Thus our approach facilitates 
the interpretation of 
observed shear fields, as we discuss further below. 

We note that the analysis presented here applies to non-singular mass
distributions which generate a gravitational potential that is both continuous and
finite everywhere, so that the induced surface is smooth. In the
case of either a point mass or an isothermal mass distribution, the
lensing potential diverges at the origin and would require further refined techniques. We expect that realistic mass profiles
and the ellipsoidal nature of halos avoids such a divergence. 

\subsection{Complex defects patterns in realistic shear maps}

The geometric analysis developed in this work can be applied to 
shear maps generated by more realistic mass distributions than the
isolated halo considered in the previous section. For instance, tidal
effects between nearby  
mass distributions can generate complex defect patterns that capture
the skeleton of the contour lines of the corresponding  
gravitational potential. In order to demonstrate this point, we have
generated shear lensing maps by means of N-body simulations combined
with a ray tracing algorithm \cite{Jain:1999ir,Stabenau}.  

\begin{figure}
\includegraphics[width=0.6\textwidth]{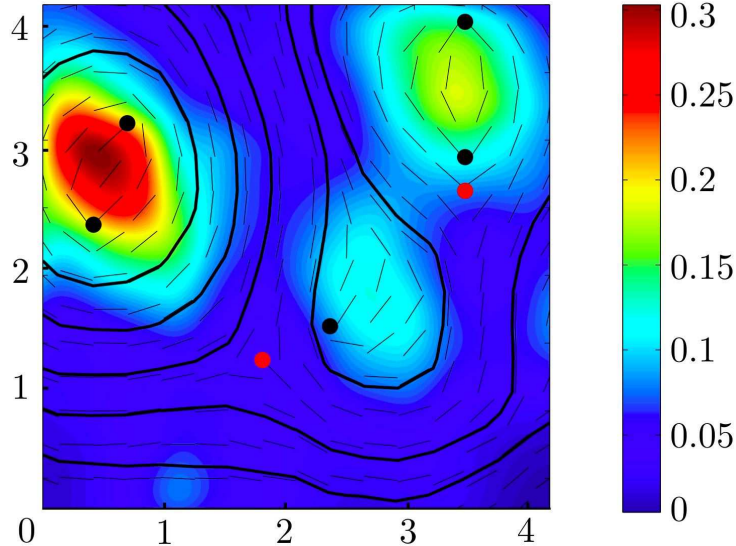}
\caption{\label{fig1} Shear lensing map generated from ray tracing
simulations on a patch of the sky with angular size around $4$ arcminutes. The wiskers represent the local direction of the shear
field interpolated by the continuous line. 
The colour plot is the projected mass distribution and the defects of
index plus or minus $1/2$ are indicated by black and red dots
respectively. } 
\end{figure}

Figure \ref{fig1} shows a patch of the observed sky with the simulated (normalized)
shear field plotted as whiskers at each pixel with the overlaid
color map of the projected mass distribution (contours of $\kappa$). 
The continuous line interpolates the whiskers to facilitate comparison
with the defect 
patterns reproduced in Fig. \ref{fig:defects}. The red and black dots
indicate negative or positive defects whose index has magnitude $1/2$
(top row in Fig. \ref{fig:defects}). They have been identified by
measuring the net change, $\Delta \alpha$, accumulated by the shear
angle in going around each pixel, see Ref. \cite{Huterer}. 

The top left part of the plot shows a prominent mass over-density of
irregular shape which generates at large scales shear field lines
topologically equivalent to a vortex. Upon magnification, two $+1/2$
defects can be resolved. The line along which the two defects lie is 
approximately oriented perpendicularly to the major axis of the nearly elliptical mass distribution, as 
indicated by our geometric analysis in the previous section.

In addition, note the presence  
of the isolated $-1/2$ defect (red dot) generated by tidal effects at
the center of three mass over densities. The local projected gravitational
potential resembles a monkey saddle when plotted as a height function
with an umbilical point (the -1/2 defect)    
at its center. Note, in addition, that the dipole of $\pm 1/2$ 
defects in close proximity
(on the right side of the picture) 
has no significant effect on the field lines at large scales
since the two defects cancel each other. 

\section{Stochastic geometry and mass fluctuations}

In this section, we consider the case of a random mass distribution and 
present a field theory from which the density
of topological defects can be read off from the statistics of the stationary random function $\psi(x,y)$. The gravitational potential is first assumed to be
a Gaussian random variable (deviations from Gaussianity can be calculated 
using perturbation theory, this is beyond the scope of this paper). 
Light coming from very distant sources undergoes deflections from a 
number of intervening lensing planes. 
Assuming that the mass distribution in different redshift slices is uncorrelated, the projected gravitational potential resulting from a large number of such lensing planes is well described by a Gaussian random variable according to the central limit theorem. In addition, with a large smoothing scale, the 
lensing mass distribution itself is well approximated by a Gaussian random
field. In practice, there are not sufficiently many uncorrelated lens planes
and the Gaussian assumption is valid only on angular scales sufficiently 
larger than 10 arcminutes (it is a better approximation for sources at higher
redshift). 

\subsection{Defect density}

In order to count $N$, the number of topological defects 
(where the magintude of $\gamma=0$) in a patch of the sky, we need to 
evaluate the surface integral
\begin{equation}
N = \int \int d \theta_1 \!\!\!\! \quad d \theta_2 \!\!\!\! \quad \delta_D(\psi_{11}-\psi_{22}) \!\!\!\! \quad \delta_D(\psi_{12}) \!\!\!\! \quad \left|\det \frac{\partial(\psi_{11}-\psi_{22},\psi_{12})}{\partial (\theta_1, \theta_2)} \right|
\label{eq:8}
\end{equation}
where $\delta_D$ indicates a delta function and the appropriate Jacobian determinant for the change of variable has been inserted \cite{Berry:1977}.
This integral can be performed explicitly once the gravitational potential $\psi(\theta_1, \theta_2)$ and its derivatives are specified. Note that the third spatial derivatives are also
needed to evaluate the determinant in Eq. (\ref{eq:8}).  

In the case of Gaussian fluctuations of the lensing potential, the {\it average} defect density can be obtained by performing a functional integral that averages over the unknown fields
with a probability density which is simply given by the exponential of a quadratic function of $\psi$ and its derivatives. Following the seminal work of Berry and Hannay \cite{Berry:1977}, we can write down
explicit formulas for the density and other statistical properties of the detects using standard field theoretic manipulations.   
The statistics of $\psi(\vec{\theta})$ are completely determined
by specifying either the autocorrelation function $C(\bold{\theta})$ or its power spectrum $P(k)$ defined by
\begin{equation}
C(\theta) = \langle{\psi(\theta_0)\psi(\theta_0+ \theta )}\rangle = \int d^2 k P(k) e^{i k \cdot \theta}
\label{eq:9}
\end{equation}
The number density of defects, $d$, can then be related to the autocorrelation function of the lensing potential 
\begin{equation}
d = \left| \frac{3 C^{(6)}_{0}}{10 \pi C^{(4)}_{0}} \right|
\label{eq:10}
\end{equation}
where the derivative $C^{(6)}_{0}$ indicates the sixth spatial derivative of the correlation function $C(\theta)$ evaluated at $\theta=0$ \cite{Berry:1977,Dennis}.

This formula can also be cast in terms of the moments of the power spectrum $P(k)$ from Eq. (\ref{eq:9}). The result reads 
\begin{equation}
d = \left| \frac{M_6}{4 \pi M_4} \right|
\label{eq:11}
\end{equation}
where the $n^{th}$ moment of the power spectrum is defined as
\begin{equation}
M_n \equiv 2\pi \!\!\!\! \quad \int_{0} ^{\infty} dk \!\!\!\! \quad k^{n+1} \!\!\!\! \quad P(k) 
\label{eq:12}
\end{equation}

The two correlation functions of topological charges with equal or opposite sign, $\{g_{++}(\theta)=g_{--}(\theta),g_{+-}(\theta)\}$, can be readily obtained from Eq. (\ref{eq:8}) in terms of $\theta$, the angular distance in the projected sky between two defects of equal or opposite index. The resulting mathematical expressions are too complicated
to list here, see \cite{Dennis} and references therein for more details. The topological defects correlation functions depend on the full functional form of $C(\theta)$ and its derivatives, not only its asymptotic value for $\theta \rightarrow 0$. As a result, they are a more sensitive probe of the underlying cosmological processes than 
the defect density in Eq. (\ref{eq:10}). 
 
The mathematical techniques adopted to study the topology of the shear field
can be applied successfully to the study of the cosmic microwave
polarization field  \cite{Huterer,Vachaspati,Novikov}. The components of the lensing shear field are always
correlated because in the standard weak lensing approximation they
are derived, via Eq. (\ref{eq:3}), from the Hessian of the gravitational potential which is the fundamental 
physical field that controls the statistics. 
(Departures from this behavior are 
indicative of systematic errors in the data or physical effects distinct
from lensing. )
The CMB polarization components 
have a B-mode contribution from primordial tensor mode fluctuations
and from lensing along the line-of-sight, 
so one generally requires both a scalar and
pseudoscalar potential to describe them \cite{HuWhite}.

\subsection{Application to the cold dark matter model}

The relations in Equations (\ref{eq:10}-\ref{eq:11}) express the number density of defects $d$ in terms
of the correlation function or power spectrum of the projected
gravitational potential. We can evaluate these for the current
cosmological best fit $\Lambda$-CDM model, with the caveat that the
Gaussian approximation breaks down on scales below about $10$
arcminutes (or angular wavenumbers $\ell$ above about 1000), the precise value of the angle requires careful tests and depends 
on the source redshift. The
breakdown occurs due to nonlinear gravitational dynamics which couples
the initially random Fourier modes that comprise the perturbed potential \cite{Dodelson:2003}. While nonlinear effects are evident in measures like the lensing power spectrum on larger scales, the Gaussian approximation may still be valid to study topological features which can be altered only by orbit crossings or mergers. 

The potential is obtained from the projected density field via the
Poisson equation as discussed in \S II. 
The density field is characterized by its power
spectrum, which can be approximated analytically by 
\begin{equation}
P(k) \approx A k T^2(k)
\end{equation}
where the transfer function $T(k)$ can be approximated analytically 
in terms of $q=k/\Omega_m h^2$ (Bardeen et al 1984)
\begin{equation}
T(q) \approx \frac{\ln(1+2.34q)}{2.34q}\left[ 1 + 3.89q + (16.1q)^2 + 
 (5.46 q)^3 + (6.71 q)^4\right]^{-1/4}
\end{equation} 

This power spectrum, evolved in time using linear perturbation theory,
and projected appropriately along the line of sight gives $C(\theta)$, the two-point correlation function of
$\psi$ \cite{Jain:1999ir}. Upon evaluating Eq. (\ref{eq:10})
above, we find (for comparison with simulations, evaluating the expression
at an arcminute), the predicted number density of defects is $\sim
10^7$ per steradian. The two main approximations made are the
assumption of Gaussianity/linear evolution and the assumption that
the asymptotic behavior as $\theta$ approaches 0 is recovered at scales of
order an arcminute. 

We compare the analytical prediction made above with the measured
number of $\pm 1/2$ defects in ten independent $\sim 6$ square degree patches
obtained from simulated shear maps. We
estimate the measured density is $2 \times 10^6$. (The error estimated from the standard deviation is less than $5 \%$ .). We may expect that
non-Gaussianity, due to the mergers of several small halos into larger
ones, tends to reduce the number of defects. Hence, a difference
within a factor of order unity is perhaps not surprising.
A more careful comparison will be attempted by making perturbative 
corrections to the Gaussian formula in future work. 

\section{Conclusion}

In this paper we have studied topological defects in the shear field 
generated from weak gravitational lensing.
Our geometric approach rests on the observation that topological defects 
correspond to umbilical point of an
imaginary surface whose height function is given by the projected 
gravitational potential. This allows us, for example, to
relate the ellipticity of a gravitational halo to the distance between two 
defects of index $+1/2$. We describe the overall pattern of the shear
field in terms of the defects generated by clustered halos, as shown in 
Figure 4. Moreover the density of the defects can yield information on the 
two point correlation function of the gravitational potential if the 
fluctuations are assumed to
be Gaussian. In this case, the statistical properties of the shear field 
can be readily calculated
from a simple field theory. 

Topological defects in the shear field provide a different view than direct
measurements of shear correlations, which is the standard approach in 
lensing cosmology. Whether the study of defects can provide cosmological
information is an open question. We have not studied the robustness of 
defect identification or their statistical properties in the presence 
of measurement noise. This should be straightforward to carry out, 
at least with Gaussian noise added to simulated shear fields. 

The analytical results in Section IV rely on the statistics of a Gaussian
random field. On small scales, the lensing shear has distinct non-Gaussian
features, so the Gaussian  description is valid only on large enough scales 
or alternatively from sources 
at very high redshift. It may however be a suitable starting point for 
perturbative calculations of the effects of weak non-Gaussianities
associated with the onset of the non-linear gravitational dynamics regime. 
In general, measurements from simulations can be used to test analytical 
results, to  compare with data 
and to find signatures of non-Gaussian features in the defect statistics. 

\begin{acknowledgments}
We wish to thank M. Dennis, W. Irvine, M. Jarvis, M. Lima, T. Lubensky, and R. Sheth for stimulating
discussions. We are grateful to H. Stabenau for help with simulation
data and several useful coversations. This work is partially funded
by NSF grants AST-0607667 and DMR05-47230, the Research Corporation, and
gifts from L.J. Bernstein and H.H. Coburn. VV acknowledges financial support from grant
DOE-DE-FG02-05ER46199.
\end{acknowledgments}

\end{document}